\documentclass[aps, prd, groupedaddress, preprint, tightenlines,  eqsecnum, nofootinbib, showpacs]{revtex4}

\usepackage{epsfig}

\usepackage{float, slashed, graphicx, amssymb, amsmath}
\usepackage[usenames, dvipsnames]{xcolor}
\usepackage{booktabs}

\usepackage[margin=2.2cm, a4paper, includefoot]{geometry}

\def\npb#1#2#3{{\rm Nucl. Phys. B}{\bf \ #1}, #3 (#2)}
\def\spa#1.#2{\left\langle#1\,#2\right\rangle}
\def\spb#1.#2{\left[#1\,#2\right]}
\def\eps{\epsilon}

\def\NeqEight{\Neq8}
\def\NeqSix{\Neq6}
\def\NeqFour{\Neq4}
\def\NeqOne{\Neq1}

\def\coli#1#2{\mathop{\longrightarrow}^{#1 \parallel #2}}
\def\soli#1{\mathop{\longrightarrow}^{k_{#1} \rightarrow0}}

\newcommand{\oneloop}{\text{1-loop}}

\newcommand{\tree}{\text{tree}}
\newcommand{\Neq}[1]{\mathcal{N} = #1}

\DeclareMathOperator{\tr}{\mathrm{tr}}
\DeclareMathOperator{\SP}{\mathrm{Sp}}
\DeclareMathOperator{\Soft}{\mathrm{Soft}}

\def\SoftP{\hat S}
\def\SoftA{S}

\def\be{\begin{equation}}
\def\ee{\end{equation}}

\begin{document}

\title{The $n$-point MHV one-loop Amplitude in $\NeqFour$ Supergravity}

\author{David~C.~Dunbar, James~H.~Ettle and Warren~B.~Perkins}

\affiliation{
College of Science, \\
Swansea University, \\
Swansea, SA2 8PP, UK\\
}

\begin{abstract}
 We propose an explicit formula for the $n$-point MHV one-loop
 amplitude in a $\NeqFour$ supergravity theory. This formula is
 derived from the soft and collinear factorisations of the
 amplitude. 
\end{abstract}

\maketitle

\section{Introduction}

``Maximal-Helicity-Violating'',  or MHV, amplitudes are scattering
amplitudes where exactly two outgoing massless particles have negative
helicity and the remaining legs have positive helicity. These objects
have been key in many of the recent developments in perturbative gauge
theories.  The Parke-Taylor formulae~\cite{ParkeTaylor}
gave  simple explicit formulae for the MHV tree-level scattering
of $n$-gluons in a colour-ordered formalism.  The simplicity and
properties of this expression have even led to the extremely fruitful
suggestion the MHV amplitudes be promoted to the underlying vertices
of the theory~\cite{Cachazo:2004kj}. 
  
At one-loop level the MHV $n$-gluon amplitudes have also been
determined analytically. 
The  different particle types contributing to
the internal loop can be conveniently
organised into supersymmetric multiplets. 
The $\NeqFour$ amplitudes were constructed first~\cite{Bern:1994zx}
with the $\NeqOne$ following shortly afterwards~\cite{Bern:1994cg}.
The remaining scalar has now finally succumbed to analytic attack~\cite{Bern:1994cg,Bedford:2004nh,Berger:2006vq}.

Graviton scattering amplitudes are considerably more
computationally complex. Expressions for the $n$-point MHV tree amplitudes were
constructed from their soft and collinear
factorisations~\cite{BerGiKu} and  subsequently proven to be
correct in
ref.~\cite{Mason:2008jy}. 
We will be
following the approach of constructing rational terms from their soft
and collinear factorisations in this article.  To date, at one-loop the only known all-$n$ rational piece of
a gravity amplitude is for the pure gravity ``all-plus'' case.
The existence of a compact, all-$n$ expression for the rational
part of a less-than-maximally supersymmetric supergravity theory
is significant, as it is a manifestation of a yet-to-be understood
pattern of simplifications underpinning quantum gravity theories.
These explicit expressions are a rich source of data for resolving
this hidden structure.

 For one-loop amplitudes there are more
potential components which can be organised into ${\cal N}=8,6,4,1,0$ matter
contributions.   The $\NeqEight$ MHV contribution was first calculated in
ref~\cite{MaxCalcsB} where a remarkable similarity between these
amplitudes and those of $\NeqFour$ Yang-Mills was observed: the
one-loop amplitudes are comprised entirely of scalar box integrals
with rational coefficients. This ``no-triangle'' feature has been
shown to extend to all one-loop $\NeqEight$ amplitudes~\cite{BjerrumBohr:2006yw}.
In terms of these matter contributions the $\NeqFour$ supergravity one-loop amplitude is
\be
M^{N=4}_n= M^{N=8}_n- 4M^{N=6,matter}_n   + 2 M^{N=4,matter}_n
\ee
Extensions to the basic $\NeqFour$ theory can be obtained from
variants of this formula. 

One-loop amplitudes in gauge and gravity theories have an expansion in
terms of scalar  $n$-point integral functions $I_n$ which encompass the transcendental
functions together with a rational remainder $R_n$.  
Amplitudes involving massless particles contain scalar functions with
$n=2,3,4$,  to order $O(\eps)$ in the dimensional regularisation
parameter.
The symmetries of the specific  theory reduce the general case to
\begin{align}
M^{N=8}_n &=
\sum  a_i I_4^i  
\notag\\
M^{N=6}_n &=
\sum  a_i I_4^i  +\sum b_j I_3^j =\sum  a_i I_4^{{\rm tc}:i} +\sum b_{k} I_3^{3m:k}
\notag\\
M^{N=4}_n &=
\sum  a_i I_4^{{\rm tc}:i}  +\sum b_j  I_3^{3m:j}+ \sum c_k I_2^k +R_n
\end{align}
where
$I_4^{\rm tc}$ is the specific combination of a scalar box integral
with its descendant scalar triangles which is IR finite.  

For the MHV helicity configurations the three mass triangles
$I_3^{3m}$ are absent and the sum of boxes is restricted to
the ``two-mass-easy'' boxes and the one-mass boxes. For $\NeqEight$
all boxes contribute whereas for $\NeqSix,4$  the boxes are restricted
to the configurations where the negative helicities
appear on opposite massive legs. 
The $a_i^{N=8}$ were first computed in ref.~\cite{MaxCalcsB}.
\be
a_i^{N=8} ={(-1)^n\over 8}\spa{m_1}.{m_2}^8  h(a,M,b)h(b,N,a) \tr(aMbN)^2 
\ee
[Note the  normalisations of the physical amplitude are 
${\cal  M}^\tree=(\kappa/2)^{n-2} M^\tree, 
{\cal  M}^\oneloop=(\kappa/2)^{n} M^\oneloop$.]
The ``half-soft''
functions $h(a,M,b)$ have the explicit form given in ref.~\cite{MaxCalcsB}. 
We are using the usual spinor products (see eg. \cite{Dunbar:2011xw}).
The $a_i^{N=6,4}$ take the related form
\begin{align}
a_i^{N=6,4} &={(-1)^n \over 8}{ \spa{m_1}.{m_2}^{8-2A} \over \spa{a}.b^{2A}}
h(a,M,b)h(b,N,a) 
\notag \\
&\times \tr(aMbN)^2
(\spa{a}.{m_1}\spa{a}.{m_2}\spa{b}.{m_1}\spa{b}.{m_2} )^A
\end{align}
with $A=1$ for $\NeqSix$ and $A=2$ for $\NeqFour$.  
The sum over bubbles includes bubbles where the clusters contain
exactly one negative helicity leg and at least one positive helicity
leg.  
The explicit form of the $c_i^{N=4}$ is given in the appendix of
ref.~\cite{Dunbar:2011xw}.  The $a_i$ and $c_i$ are, in general,
computed using unitarity based techniques.  

The rational terms $R_n$ cannot be determined by four dimensional
unitarity. They may be determined using ``D-dimensional'' unitarity~\cite{Dunitarity}
but are thus more difficult to obtain.  
Explicit computations of $R_n$ have been restricted to $n=4,5$~\cite{GravityStringBasedB,Dunbar:2010fy,Bern:2011rj}. 
In the next section we present an all-$n$ form of $R_n$ thus
completing the $\NeqFour$ one-loop MHV amplitude.

\section{Result}

The $n$-point rational term, which completes the $n$-point amplitude
$M(m_1^-,m_2^-,p_1^+,\cdots p_{n-2}^+)$ is
\be
R_n 
=(-1)^n{  \spa{m_1}.{m_2}^4\over 2}  \biggl( R_n^0+   \sum_{r=3}^{n-2} R_n^r  \biggr) 
\ee
In the above
\begin{align}
R_n^0= 
\sum_{boxes}
{\spb{a}.b^2 \over \spa{a}.b^2 }& h(a,M,b)h(b,N,a) 
  \notag\\
&\times
\left(\spa{m_1}.a \spa{m_2}.a \spa{m_1}.b \spa{m_2}.b
\right)^2 
\end{align}
where there is a contribution for each box integral function present in the amplitude.
The $R_n^0$ contain spurious quadratic singularities which are
necessary to cancel those in the box integral
contributions~\cite{Dunbar:2011xw}.
The remaining $R_n^r$
are 
\be 
R_n^r = \sum_{subsets}    C_{r} [\{ p\}^{(r)}]  
\times \SoftP^{n-2-r}_{ \{ q\}^{(n-2-r)}}  
\ee
The sum is over subsets $\{p\}^{(r)}$ of $\{p_1,\cdots p_{n-2}\}$ of
length $r$ of which there are
$(n-2)!/r!/(n-2-r)!$.  The $\{q\}^{(n-2-r)}$ are the remaining positive helicity legs.
The $C_r$ are 
\be
C_r[ \{p_1,\cdots p_r\}]=
\sum_{perms}  
{ \spb{p_1}.{p_2} \spb{p_2}.{p_3} \cdots \spb{p_r}.{p_1}
\over
\spa{p_1}.{p_2} \spa{p_2}.{p_3} \cdots \spa{p_r}.{p_1} }
\ee
where the sum over permutations is over the $(r-1)!$ cyclically independent
choices of orderings of $\{p_1,\cdots p_r\}$.

The $ \SoftP^m $ are polynomial in the objects $A[a;s]$, 
\be
A[a;s]  \equiv { \spb{s}.{a}\spa{a}.{m_1}\spa{a}.{m_2}   
\over \spa{s}.{a}
  \spa{s}.{m_1}\spa{s}.{m_2} } 
\ee
and are best defined by their soft behaviour as $s\in \{q\}^m$ becomes
soft, 
\be
 \SoftP^m_{\{ q\}^m}
\longrightarrow   -\Soft(s^+) \times \SoftP^{m-1}_{\{q\}^m-s}
\ee
where $\Soft(s^+)$ is the soft-factorisation function~\cite{BerGiKu},
\be
\Soft(n^+) =-{ 1\over \spa{1}.n\spa{n}.{n-1} } \sum_{j=2}^{n-2} {
\spa{1}.j  \spa{j}.{n-1} \spb{j}.n\over \spa{j}.n } .  
\label{softfn}
\ee
together with the restriction that {\it any cyclic combinations of
  $A[q_i,q_j]$  are excluded}.  Note that the negative sign is
necessary since there is an overall factor of $(-1)^n$ in the
amplitude. The $\SoftP_m$   take the form
\be
\SoftP^m_{\{q\}^m} =\prod_{k=1}^m\SoftP^1_{q_k} - {\it cycle \; terms}
\ee

\def\jam{\hskip -10pt}
\def\jamr{\hskip -15pt}
The first few are given by, (note $\SoftP^0 =1$ )
\begin{align}
&\SoftP^1_{q_1} = \sum_{p_j\in \{p\}^{(n-3)}} \jam A[p_j;q_1]
\notag \\
&\SoftP^2_{\{q_1,q_2\}} =\sum_{p_j\in \{p\}^{(n-4)}} \jam A[p_j;q_1]
\sum_{p_l\in \{p\}^{(n-4)}} \jam A[p_l;q_2]
\notag \\ 
&\hskip 10pt
+ A[q_1;q_2]\jam\sum_{p_j\in \{p\}^{(n-4)}} \jamr A[p_j;q_1] 
+ A[q_2;q_1]\jam\sum_{p_j\in \{p\}^{(n-4)}} \jamr A[p_j;q_2]
\notag \\ 
&\hskip 20pt
= \SoftP^1_{q_1} \SoftP^1_{q_2} -A[q_1;q_2]A[q_2;q_1]
\end{align}

The cyclic combinations of $A[q_i;q_j]$ simplify into cyclic
combinations of 
$\spb{q_i}.{q_j}/\spa{q_i}.{q_j}$, e.g. 
\be
A[q_1;q_2]A[q_2;q_1] = { \spb{q_1}.{q_2}^2 \over \spa{q_1}.{q_2}^2 }
\ee
and are non-singular in the soft limit.   We will present an alternative
description of $R_n$ in the next section.

The structure of $R_n$ is a rational function of the spinor variables
$\lambda_{a}^i$ and $\bar\lambda_{\dot a}^i$. The function is rational
in $\lambda_a^i$ but only polynomial in $\bar\lambda_{\dot a}^i$ the
polynomial being homogeneous of degree $2(n-2)$.  The tree MHV
amplitude shares this feature but the polynomial is of degree
$2(n-3)$.   Consequently the
$R_n$ have an analogous ``twistor-space'' structure to the MHV tree
amplitudes~\cite{Witten:2003nn,Bern:2005bb}.  

\section{Construction}

The form of $R_n$ was obtained from soft and collinear
factorisations.  Note that an MHV amplitude in a supergravity theory
does not have any physical multi-particle poles. 
The collinear limit occurs when legs $k_a$ and $k_b$ are collinear,
$k_a\cdot k_b \longrightarrow 0$.  Unlike Yang-Mills
amplitudes, gravity amplitudes are not
singular in the collinear limit, but acquire a ``phase-singularity''~\cite{MaxCalcsB}  that is specified in
terms of amplitudes with one less external leg. 
If
$k_a \longrightarrow z K$ and $k_b \longrightarrow (1-z) K $,
\be
M_n( \cdots , a^{h_a}, b^{h_b} ) \coli{a}{b}   \sum_{h'}
\SP_{-h'}^{h_ah_b}  M_{n-1} (\cdots , K^{h'} )   +F_n
\ee
where the $h$'s denote the various helicities of the
gravitons and $F_n$ indicates the remainder term with no  phase
singularity. 
The non-zero
``splitting functions'' are~\cite{MaxCalcsB}
\begin{align}
   \SP_{-}^{++}&= -{ \spb{a}.{b} \over z(1-z)  \spa{a}.{b}  }, 
\;\;\; 
  \SP_{+}^{-+} 
= -{ z^3\spb{a}.{b} \over (1-z) \spa{a}.{b} }.
\end{align}
Gravity amplitudes also have soft-limit singularities~\cite{BerGiKu}
as $k_n \longrightarrow 0$,
\be
M_n( \cdots , n-1, n^h) \soli{n}  \Soft (n^h) M_{n-1} (\cdots ,
n-1 ) .
\ee
An important result of ref.~\cite{MaxCalcsB} is that the splitting and
soft factorisation functions \eqref{softfn} do not obtain loop corrections. 
The entire amplitude must satisfy these soft and collinear
factorisations.  With the exception of the collinear limit of two
positive helicity legs,  the transcendental functions and rational
term  factorise independently~\cite{Dunbar:2011xw,Dunbar:2010fy}.

When considering these limits it is useful to use an alternate form: 
\be
\sum_{r=3}^{n-2} R_n^r =\sum_{r=3}^{n-2}\sum_{subsets}  \hat  C_r  [\{ p_j\}] 
  \SoftA^{n-r-2}_{\{ q\}^{(n-r-2)}}  
\ee
where
\be
 \SoftA^{m}_{\{ q\}^m} 
=\prod_{k=1}^m \hat \SoftA^1_{q_k}
\ee
and  
\be
 \hat C_r  [ \{ p \} ] = C_r  [ \{ p \} ] -\sum_{s=2}^{r-2} \sum_{subsets}  \epsilon_{r,s} C_{r-s} [\{ b\}]    
\times  C_{s}[\{c\}]+\cdots 
\ee
where $\{c\}$ is a subset of $\{p\}$ of length $s$ and $\{b\}=\{p\}-\{c\}$.   
The $\hat C_r$ is simply the weighted sum of all single and multiple
cycles.  $\hat C_3=C_3$ and $\hat C_4=C_4$ but
\begin{align}
\hat C_5[\{a_1\cdots & a_5\}]
= C_5 [\{a_1\cdots a_5\} ] 
\notag\\- \sum_{subsets} & C_2[ \{a_1,a_2\}]]  C_3[\{a_3,a_4,a_5\}]
\end{align}
or using simplified notation,
\begin{align}
\hat C_5 &=C_5-C_2C_3
\notag\\
\hat C_6
&= C_6 -C_2C_4- C_3C_3  
\notag\\
\hat C_7
&= C_7 -C_2C_5-2C_3C_4+C_2C_2C_3/2
\notag\\
\hat C_8 &=C_8-C_2C_6-2C_3C_5-C_4C_4+C_2C_3C_3
\notag\\
& \hskip 25pt+C_2C_2C_4/2 
\end{align}
In this form, the cycle terms previously subtracted from $\SoftP^m$
lie with the $\hat C_r$ terms, leaving $\SoftA^m$ which have simpler
soft and collinear behaviour.
This form is useful in examining the soft and collinear limit but is
really a more complicated expression where material has been added to
both $\SoftP_r$ and $C_r$.

 The soft-behaviour of $R_n^0$  can be derived from the soft-behaviour  of the
half-soft functions~\cite{MaxCalcsB},   
\be
h(a,M,b) \soli{m}  -\Soft_m(a,M,b) h(a,M-m,b)  
\ee
for  legs $m\in M$. Where 
\be
\Soft_m(a,M,b)=
-{ 1\over \spa{a}.m\spa{m}.{b} } \sum_{j\in M-m} {
\spa{a}.j  \spa{j}.{b} \spb{j}.m\over \spa{j}.m } .  
\ee
From this property of the half-soft functions we can show
\be
R_n^0 \soli{n}  -\Soft(n^+) \times R_{n-1}^0
\ee

The soft-behaviour of $\SoftA_1$ is quite clear
\be
\SoftA^1_q   
\soli{q} -\Soft(q^+)
\times 1, \quad
\SoftA^1_q 
\soli{other}
 {\rm finite} 
\ee
so
\begin{align}
\SoftA^m_{\{ q\}^m}  &\soli{s}  -\Soft(s^+) \times
\SoftA^{m-1}_{\{ q\}^m-s}   \;\;\;\; s\in \{  q \}^m 
\end{align} 
The $\hat C_r$ do not contribute to any soft-singularity    
so we can deduce 
\be
R_n^i \soli{p}  -\Soft(p^+) \times R_{n-1}^i ,\quad i=3,\dots, n-3.
\ee
The term $R_n^{n-2}$ has no soft singularity.

\noindent
{\it Collinear Limits} 
There are three types of collinear limit. The amplitudes
vanishes as 
two negative legs become collinear as we would expect since the 
$\it daughter$ amplitude
$M(K^-,3^+,\cdots, n^+)$ vanishes in any supersymmetric
theory.  There are two non-vanishing independent collinear limits -
where the legs are $(m^-,p^+)$ and $(p_a^+,p_b^+)$.   (Multi-collinear
limits for this case give no further constraints beyond iteratively
applying two-particle collinear limits.)

First consider the limit
$(m^-,p^+)$.  Note that $R^{n-2}_n$ does not contribute to this limit. 
The function $S^1_q$ has no collinear
phase singularity unless $q=p$,  in which case
\begin{align}
S^1_{p}  \coli{m}{p}
{ z^2 \spb{p}.{m}\over  z(1-z)   \spa{p}.{m} } 
=- { 1\over z^2 } \SP^{-+}_+ 
\label{s1splitEQ}
\end{align}
So
\be
S^t_{\{q\}^t}  \coli{m}{p}
- { 1\over z^2 } \SP^{-+}_+ 
\times S^{t-1}_{\{q\}^t-p} 
\ee
for $p \in  \{q\}^t$ and zero otherwise. 
The $\hat C_r$ do not contribute to this collinear limit and so we 
deduce
\be
\spa{m}.{m'}^4 R_n^i  
\coli{m}{p}  -\SP^{-+}_+ \spa{K}.{m'}^4  R_{n-1}^i
\ee
with the factor of $z^2$ from $\spa{m}.{m'}^4$ cancelling the
$z^{-2}$ in (\ref{s1splitEQ}). 
The collinear limit of $R_n^0$ follows from the 
collinear behaviour~\cite{MaxCalcsB} 
of the half-soft functions
\be
h(a,\{c,d,\cdots  \},b) \coli{c}{d}
{ 1\over z(1-z) } { \spb{c}.d\over \spa{c}.d } h(a,\{K,\cdots\},b)
\ee 
from which we can deduce, 
\be
\spa{m}.{m'}^4 R_n^0 \coli{m}{p}
-\SP^{-+}_+  \times \spa{K}.{m'}^4 R_{n-1}^0 
\ee

The $(p_a^+,p_b^+)$ collinear limit is a little more subtle.  The
terms in $R^0_n$ with 
a double phase singularity $\sim { \spb{a}.b^2 /\spa{a}.{b}^2 }$
cancel exactly against the corresponding box integral contributions as
$s_{ab} \longrightarrow 0$ and give no phase singularity.  
The remaining terms  in $R^0_n$ we refer to as $R^0_n|_{red}$ and should satisfy:
\be
R^0_n|_{red} +\sum_{i=3}^{n-2}  R^i_n \longrightarrow \SP^{++}_-  \times
\left( R_{n-1}^0+
\sum_{i=3}^{n-3}  R^i_{n-1}.
\right)
\ee
Note that this is the only factorisation the term 
$R^{n-2}_n$  contributes to. 
Although, at present, we have  no analytic proof that the $n$-point expression has
the correct collinear limit we have checked this numerically up to
ten-points.  Note that, unlike the $(m^-,p^+)$ collinear
limit it is not satisfied  ``term-by-term''  for the $R^i_{n}$
but only by the total. 

The expression for $R_n$ gives a candidate amplitude which
satisfies all physical collinear and soft-factorisations, contains no
spurious singularities and satisfies the expected symmetries of the
amplitude.  We do not possess a proof that this
expression is correct beyond five-points although experience suggests it is extremely
likely to be so: for the MHV tree amplitudes~\cite{BerGiKu},
the $\NeqEight$ MHV and the all-plus one-loop
amplitudes~\cite{MaxCalcsB}, soft and collinear constraints
were sufficient to generate expressions which were subsequently
proven correct. Indeed, there are recent
suggestions~\cite{Nguyen:2009jk} that soft-limits alone may
determine tree amplitudes.

\section{Conclusions}

At present the perturbative
structure of (super)gravity theories appears to be considerably more
constrained with hidden structures and more symmetries than were apparent only a
few years ago.
The existence of explicit amplitudes is of key
importance in forming and testing conjectures in perturbative field
theory.
 Currently, very few explicit loop amplitudes exist to
test perturbation theory beyond tree level in gravity theories. 
We have proposed an expression for the $n$-graviton MHV one-loop
amplitude in $\NeqFour$ supergravity.  This
expression adds to a very small list of all-$n$
one-loop expressions in gravity :  the $\NeqEight$ and $\NeqSix$ MHV
amplitudes and the pure gravity ``all-plus'' amplitude.  
Such explicit expressions have
been extremely useful in the past in elucidating the perturbative structure of
gauge theories.  Our  expression provides a goal for other approaches
such as for example the gauge-gravity  
conjectures~\cite{Bern:2011rj,BoucherVeronneau:2011qv,Bern:2010yg}. 
In general,  we hope that this series
of amplitudes will prove useful in untangling the perturbative expansion
of quantum (super)gravity.

\end{document}